\documentclass[aoas,nameyear,dvips]{arximspdf}

\doi{10.1214/10-AOAS365}
\volume{4}
\issue{2}
\pubyear{2010}
\firstpage{533}
\lastpage{534}

\begin{document}
\begin{frontmatter}

\title{Introduction to papers on the modeling and analysis of network data---II}
\runtitle{Section on Network Modeling---II}

\begin{aug}
\author{\fnms{Stephen E.} \snm{Fienberg}\ead[label=e1]{fienberg@stat.cmu.edu}\corref{}}%\thanksref{t1}}
\runauthor{S. E. Fienberg}
\address{Department of Statistics\\
\quad and Machine Learning Department\\
Carnegie Mellon University\\
Pittsburgh, Pennsylvania 15213\\
USA\\
\printead{e1}}
\affiliation{Carnegie Mellon University} %adresu isvedimo komanda gale!
\end{aug}

% HISTORY:
\received{\smonth{5} \syear{2010}}

% ABSTRACT

% KEYWORDS

\end{frontmatter}

This issue of \textit{The Annals of Applied Statistics} (Volume 4, No. 2)
contains the second part of a Special Section on the topic of network
modeling. The first part consisted of seven papers and appeared with a
general introduction [Fienberg~(\citeyear{Fien2010})] in Volume 4, No. 1. In Part II we
include a diverse collection of eight additional papers with
applications spanning biological, informational and social networks,
using techniques such as kriging and anomaly detection, and variational
approximations, as well as the study of latent structure in both static
and dynamical networks:

\begin{itemize}
\item In \textit{A State-Space Mixed Membership Blockmodel for Dynamic
Network Tomography}, Xing, Fu and Song combine earlier approaches
involving mixed membership stochastic
blockmodels for static networks with state-space models for
trajectories and use the new dynamic modeling approach to analyze the
Sampson's network of noviates in a monastery, the email communication
network between the Enron employees and a rewiring gene interaction
network of the life cycle of the fruit fly.

\item In \textit{Maximum Likelihood Estimation for Social Network
Dynamics}, Snijders, Koskinen and Schweinberger develop a
likelihood-based approach to network panel data with an underlying
Markov continuous-time stochastic actor-oriented process. They use the
new methods to reanalyze a friendship network
between 32 freshman students in a given discipline at a Dutch
university, observed over six waves at three-week intervals beginning
at the start of the academic year.

\item Xu, Dyer and Owen use a semi-supervised learning on network
graphs in which response variables observed at one node are used to
estimate missing values at other nodes, by exploiting an underlying
correlation structure among nearby nodes. The methods they employ in
\textit{Empirical Stationary Correlations for
Semi-supervised Learning on Graphs} are rooted in ideas about kriging
emanating from geostatistics, and they compare their methods to ones
proposed earlier using a data set containing the number of web links
between UK universities in 2002, and the WebKB data set containing
webpages collected from computer science departments
of various US universities in 1997.

\item In \textit{Ranking Relations Using Analogies in Biological and
Information Networks}, Silva, Heller, Ghahramani and Airoldi explore
the problem of ranking relations in network-like settings based on a
similarity criterion underlying Bayesian sets, drawing on ideas of
analogy items in test batteries such at the SAT. They too analyze the
WebKB collection, as well as the problem of ranking protein--protein
interactions using the MIPS database for the proteins in budding yeast.

\item Heard, Weston, Platanioti and Hand fuse discrete time counting
models to carry out \textit{Bayesian Anomaly Detection Methods for Social
Networks} using data from the European Commission Joint Research Centre's
European Media Monitor web intelligence service, that provides
real-time press and media summaries
to Commission cabinets and services, including a breaking news and
alerting service. They also study simulated cell phone data from the
VAST Mini Challenge covering a fictional ten-day period on an island,
narrowed to 400 unique cell phones during this period.

\item James, Zhou, Zhu and Sabatti study \textit{Sparse Regulation
Networks}, in genetic contexts using prior information about the
network structure in conjunction with observed gene expression data to
estimate the transcription regulatory network for \textit{E. coli}. Their
approach uses $L_1$ penalties on the network to ensure a sparse structure.

\item Zanghi, Picard, Miele and Ambroise explore \textit{Strategies for
Online Inference of Model-Based Clustering in Large and Growing
Networks}. Their online EM-based algorithms offer a good trade-off
between precision and speed, when estimating parameters for mixture
distributions applied to data from the political websphere during the
2008 US political campaign.

\item Mariadassou, Robin and Vacher, in \textit{Uncovering Latent
Structure in Valued Graphs: A Variational Approach}, use variational
approximations to likelihood mixture modes where the network
connections are weighted values instead of simple 0--1 entries. They use
their method to analyze interaction networks of tree and fungal species.

\end{itemize}
\def\bibname{Reference}

\printaddresses

\end{document}